# Non-Fermi liquid behavior in a mixed valent metallic pyrochlore iridate $Pb_2Ir_2O_{7-\delta}$

Md Salman Khan[1], Ilaria Carlomagno[2], Carlo Meneghini[3], P. K. Biswas[4], Fabrice Bert[5], Subham Majumdar[6], and Sugata Ray[1,7*]

[1] *School of Materials Science, Indian Association for the Cultivation of Science, 2A & 2B Raja S. C. Mullick Road, Jadavpur, Kolkata 700032, India*

[2] *Elettra Sincrotrone Trieste, Area Science Park 34149, Basovizza (TS), Italy*

[3] *Dipartimento di Scienze, Universitá Roma Tre, Via della Vasca Navale, 84 I-00146 Roma, Italy*

[4] *ISIS Facility, Rutherford Appleton Laboratory, Chilton, Didcot, Oxon OX110QX, United Kingdom*

[5] *Université Paris-Saclay, CNRS, Laboratoire de Physique des Solides, 91405, Orsay, France*

[6] *School of Physical Science, Indian Association for the Cultivation of Science, 2A & 2B Raja S. C. Mullick Road, Jadavpur, Kolkata 700032, India and*

[7] *Technical Research Center, Indian Association for the Cultivation of Science, 2A & 2B Raja S. C. Mullick Road, Jadavpur, Kolkata 700032, India*

(Dated: December 30, 2021)

## Abstract

Non-Fermi liquid behavior in some fermionic systems have attracted significant interest in last few decades. Certain pyrochlore iridates with stronger spin-orbit coupling strength have recently been added to the list. Here, we provide evidence of such a non-Fermi liquid ground state in another mixed valent metallic pyrochlore iridate $Pb_2Ir_2O_{7-\delta}$, through the combined investigation of electronic, magnetic and thermodynamic properties as a function of temperature ($T$) and applied magnetic field ($H$). Resistivity measurement showed a linear temperature dependence down to 15 K below which it shows $\rho \sim T^{3/2}$ dependence while magnetic susceptibility diverges as $\chi(T) \sim T^{-\alpha}$ ($\alpha < 1$) below 10 K. While a strong negative $\Theta_{CW}$ has been observed from Curie-Weiss fitting, absence of any long-range order down to 80 mK only indicates presence of strong inherent geometric frustration in the system. Heat capacity data showed $C_p \sim T \ln(T_0/T) + \beta T^3$ dependence below 15 K down to 1.8 K. More importantly spin-orbit coupling strength by x-ray absorption spectroscopy was found to be weaker in $Pb_2Ir_2O_{7-\delta}$ compared to other pyrochlore iridates. In absence of any large moment rare earth magnetic ion, $Pb_2Ir_2O_{7-\delta}$ presents a rare example of an irirdate system showing non-Fermi liquid behavior due to disordered distribution of $Ir^{4+}$ and $Ir^{5+}$ having markedly different strengths of spin-orbit coupling which might offer a prescription for achieving new non-Fermi liquid systems.

---

* mssr@iacs.res.in

## I. INTRODUCTION

One of the puzzling phenomena of correlated electronic materials is the understanding of the non-Fermi liquid (NFL) behavior of interacting fermions which goes beyond the Landau's Fermi liquid paradigm. Although simple metals could be aptly described by the Fermi liquid theory (FL), there are several fermionic systems such as spin-orbit assisted Mott insulator[1,2], heavy fermionic systems[3–5], one dimensional Luttinger liquids[6], and also the high $T_c$ cuprate superconductors[7,8], that cannot be dealt with the same and they are identified as the signature NFL systems. It is now understood that a finite density of fermions directly coupled to a gapless bosonic mode, as is expected near quantum critical points of metals, or two-channel and disordered Kondo systems, or spin-charge separation and the appearance of spinon and holon quasiparticles as observed in 1-dimensional Luttinger liquids, are some of the phenomena which can trigger the NFL state[9–14]. On the other hand, disorder induced NFL treated by disordered Hubbard model away from weak disorder limit[15–17] opens up another possibility, while recent treatment of random couplings in flavor space between the fermions and the bosonic order parameters without spatial randomness provides another avenue[18]. The presence of strong spin-orbit coupling (SOC) and its effect on the NFL state has also been studied by a model consisting of the Luttinger Hamiltonian supplemented by Coulomb interactions and it was observed that a quantum critical non-Fermi liquid phase can be induced provided time-reversal and cubic symmetries are maintained[19]. In this context, pyrochlore iridates with sizeable SOC have largely taken the center stage and systems like $A_2Ir_2O_7$ ($A$ = Pr, Eu) have indeed shown unusual NFL behaviors [19,20].

Experimentally it is possible to identify the NFL systems by probing the transport, thermodynamic, and magnetic behaviors, due to their distinct deviation from standard FL behavior. For example, some of the typical hallmark signature of NFL are: deviation from quadratic temperature dependence of resistivity, $\rho \sim T^n (1 \leq n \leq 2)$, logarithmic dependence in low temperature heat capacity, power law ($\chi \sim T^{-p}$; $p <1$) or logarithmic dependence of magnetic susceptibility $\chi(T)$, *etc.* And iridates, possessing a complex interplay between intra-atomic coulomb repulsion energy ($U$), noncubic crystal field energy ($\Delta^{CFE}$), intra-site hopping ($t$) and most importantly spin-orbit coupling ($\lambda$)[1,21–23] within their electronic structure, could gather maximum interest in this context, due to the experimental observation of such deviations, especially in certain members of the pyrochlore family[19,20,24].

Recently $Pr_2Ir_2O_7$ the metallic member of the pyrochlore family has been found to exhibit non-Fermi liquid behavior due to three dimensional quadratic band touching at the Brillouin zone center[19]. However, a general understanding of such NFL state in pyrochlore iridate, is still far from complete. Recently emergence of NFL state in another pyrochlore iridate, upon hole doping, has also been demonstrated in $Eu_{2−x}Sr_xIr_2O_6$[20]. The substitution of trivalent Eu by divalent Sr also leads to the suppression of long range magnetic order with simultaneous extension of metallicity down to atleast 2 K. This NFL behavior was argued to have arisen due to the disordered distribution of significantly different SOC strengths ($\lambda$)of $Ir^{4+}$ and $Ir^{5+}$, which are generated due to hole doping. It has been mentioned earlier that disorder can indeed lead to non-Fermi liquid scenario in correlated systems[13,15–18] and even induce metal to insulator transition[25], and therefore, mixed valent higher transition metals with strong SOC may be considered as a good starting point to investigate the breakdown of Fermi liquid description.

Here we report NFL-like behavior in $Pb_2Ir_2O_{7-\delta}$, another less explored member of the metallic iridate pyrochlore family. In this case, the absence of a rare earth magnetic ion at the *A*-site provides a good opportunity to independently study the physical properties associated with only Ir in this metallic sample. On the other hand, Ir is supposed to be in a $5d^4$ ($Ir^{5+}$) state in this system with comparatively weaker SOC strength than the pentavalent iridates as the SOC is renormalized in a solid as $\lambda/2S$. However, this unique system accommodates a high degree of oxygen vacancy which gives rise to a mixture of $Ir^{4+}$ ($d^5$) and $Ir^{5+}$ ($d^4$) ions even without any intentional electron or hole doping. Naturally, small changes in the structure may drastically influence the magnetic and electronic ground state properties of this compound as $Ir^{4+}$ is magnetic, whereas $Ir^{5+}$ is expected to stabilize in a nonmagnetic $J_{eff}$ = 0 ground state within the *jj*-coupling description. We found that the $x$ coordinate of the oxygen atom in $Pb_2Ir_2O_{7-\delta}$ is 0.325 ($x_c$ should be 0.3125 for ideal non distorted octahedra) which leads the Ir-O-Ir bond angle to be 134° ,which is the highest among the pyrochlore iridates at room temperature. As a consequence, the system is expected to possess large bandwidth due to enhanced effective hybridization and a moderate *U* + SOC may not be adequate to open a gap, thereby keeping the system metallic down to at least 2 K. Interestingly, the temperature variation of resistivity deviates from conventional quadratic ($\rho \sim T^2$) dependence. Our combined investigation of transport, magnetic and thermodynamic properties of the system revealed a robust non-Fermi liquid behavior without any long range magnetic order down to 80 mK as confirmed by $\mu$SR experiment.

## II. EXPERIMENTAL TECHNIQUE

Polycrystalline samples of $Pb_2Ir_2O_{7-\delta}$ were prepared by a conventional solid state reaction method using high purity (<99.9%) starting materials PbO, $IrO_2$ (from Sigma Aldrich) in proper ratio. These mixtures were thoroughly ground and pressed into pellets before initial calcination at 800°C for 12 h. Finally, the as-calcined pellet was annealed at 900°C three times for 12 h each in the oxygen atmosphere with intermediate grindings. The samples were annealed in oxygen atmosphere to avoid any vacancy due to external factors. The phase purity of the sample was checked from x-ray powder diffraction measured at Rigaku Smart Lab x-ray diffractometer with Cu-$K_\alpha$ radiation at room temperature as well as low temperature down to 4 K. The crystal structure of this sample was obtained after refining the x-ray diffraction (XRD) data by Rietveld technique using FULLPROF program[26]. Temperature dependent electrical resistivity at zero field and 5 T field was measured using the four-probe technique in Quantum Design PPMS. The x-ray photoelectron spectroscopy (XPS) measurements were carried out using OMICRON electron spectrometer, equipped SCIENTA OMICRON SPHERA analyzer and Al K monochromatic source with an energy resolution of 0.5 eV. Before collecting the spectra, the surface of the pelletized sample was cleaned in-situ by argon sputtering. The collected spectra were then processed and analyzed with KOLXPD program. The Ir $L_3$ and $L_2$ edge (~11.2 keV) x-ray absorption spectroscopy (XAS) experiment at ambient temperature was performed at XAFS beamline of Elettra synchrotron radiation facility in Italy [27]. The incident energy was set using Si (111) double crystal monochromator where couple of mirrors are used for efficient harmonic suppression. The collected XAS data were processed and analyzed using freely available DEMETER package[28,29] (ATHENA) and FITYK[30] software. Temperature and magnetic field dependent dc magnetization were carried out using the Quantum Design (SQUID) magnetometer. Heat capacity in zero field was measured using the heat capacity attachment of a Quantum Design PPMS. Muon spin resonance ($\mu$SR) experiments were performed using the MUSR spectrometer at the ISIS facility in UK.

## III. RESULTS AND DISCUSSIONS

X-ray diffraction (XRD) pattern obtained from polycrystalline $Pb_2Ir_2O_{7-\delta}$ at different temperatures are refined by Rietveld method, confirming pure single phase with cubic $Fd\bar{3}m$ space group throughout the temperature range (4-300 K). Best fitted Rietveld refined curve and the collected XRD spectrum at 4 K and at 300 K is shown in Fig. 1a. The temperature variation of cubic lattice parameter, *a*, is shown in the inset of Fig. 1b. However, no anomaly in lattice parameters and/or lifting of crystal symmetry are found which is consistent with the previous temperature dependent powder neutron diffraction study of M. Retuerto *et. al.*[31]. The detailed crystal structure parameters are given in Table-1. The refined crystal structure at room temperature is shown in Fig. 2a. The structure shows that the Ir atoms form a network of corner shared $IrO_6$ octahedra while Ir and Pb also form corner shared tetrahedral network individually (not shown). The structure also forms edge shared tetrahedral network between O´ $Pb_4$ and $OPb_2Ir_2$ tetrahedra (Fig. 2b), *i.e.*, all of this triangular network contributes to strong inherent geometric frustration in this system. The $IrO_6$ octahedra are trigonally distorted (see Fig. 2c) in the structure. Interestingly, the system happens to have a mixture of $Ir^{4+}$ and $Ir^{5+}$ states due to the tolerance of vacancy at the O´ site. To avoid any oxygen off-stoichiometry due to lack of oxygen during synthesis we annealed the sample in $O_2$ flow at 900 °C. Nonetheless, our XRD refinement confirms vacancy at O´ site in the sample, which proves that the presence of oxygen vacancy is unavoidable in this system. The stereochemically active Pb $6s^2$ lone pair remains directed towards the O´ ion within the O´ $Pb_4$ tetrahedral network through strong hybridization with O $2p$ orbital[32] which pushes the O´ ion, occupying the $4a$ site, equally from four directions (see Fig. 2e)[33]. Therefore, to energetically stabilize itself, the system tolerates some vacancy at the O´ site which leads to the formation of the mixture of $Ir^{4+}$ and $Ir^{5+}$ in the system. Charge state in all iridate systems is of central importance as the higher charge state $Ir^{5+}$ carries no magnetic moment and should possess a $J_{eff}$ = 0 state in the strong spin-orbit coupled limit (*jj*-coupling) or a magnetic state in the $L-S$ coupling limit depending on the strength of SOC, whereas $Ir^{4+}$ is magnetic and found to exhibit spin-orbit coupled $J_{eff}$ = 1/2 state commonly reported in insulating iridates. To verify the possible mixed valency of Ir, photoelectron spectra of Ir $4f$ core level along with O $1s$ level spectra were collected. The distinctively asymmetric shape of doublet Ir $4f$ core level spectrum (Fig. 3a) clearly reveals presence of two oxidation states of Ir. The spectrum could be fitted with two doublets having spin-orbit separation of 3.05 eV and 3.17 eV which confirms the presence of both $Ir^{4+}$ and $Ir^{5+}$ in the system, consistent with previous findings[34,35]. The contribution of $Ir^{4+}$ is approximately twice the contribution of $Ir^{5+}$ in the system but the ratios could also be an artifact and influenced by the well-known XPS surface effect. The oxygen 1s spectra (Fig. 3b) can also be clearly seen to consist of two distinct oxygen sites, corresponding to O 1s (529.45 eV) with higher intensity and possibly the O´ 1s level (531.14 eV) having lower intensity.

Although there are numerous 4d itinerant electron systems exhibiting intriguing quantum phenomenon, *e.g.*, p-wave superconductivity in $Sr_2RuO_4$[36], a field-tuned electronic phase in $Sr_3Ru_2O_7$[37], itinerant ferromagnetism in $SrRuO_3$[38] and bad metallicity observed in $CaRuO_3$[39], it is rare to find itinerant characteristic, although expected, among 5d iridates. Most of the iridate systems are found to be insulating due to the complex cooperation between strong spin-orbit coupling and electron-electron correlation ($U$)[1,40,41], which work together to split the wide 5d band and open a gap (left panel of Fig. 4c). Interestingly, there have been few examples which behave otherwise. The presence of finite density of states at the Fermi level at room temperature in the XPS valence band spectrum, as seen in Fig. 4a, confirms that $Pb_2Ir_2O_{7-\delta}$ also falls into that rare category. This metallic nature warrants investigation about the strength of $\lambda$ in the system. A measure of strength of SOC might give some idea about the bandwidth of the 5d level and also about the magnetism of the system. In order to quantify the same, x-ray absorption spectroscopy (XAS) data were collected at the Ir $L_2$ ($2p_{1/2} \rightarrow 5d$) and $L_3$ ($2p_{3/2} \rightarrow 5d$) edges (see Fig. 4b). By calculating the relative enhancement of $L_3$ edge compared to $L_2$ edge it becomes possible to comment on the strength of spin-orbit coupling of a system[42]. This quantification is defined by branching ratio, represented as $BR = Ir_{L3}/Ir_{L2}$. In a single particle picture, without the effect of SOC, the $J_{eff} = 5/2$ and $J_{eff} = 3/2$ states of Ir 5d valence state should be degenerate. Consequently, the transition probability to the 5d valence state from core 2p level will only give rise to the expected statistical branching ratio which is $Ir_{L3}/Ir_{L2} = 2$. Fig. 4b shows the $Ir_{L3}$ and $Ir_{L2}$ white line spectra with the corresponding fittings yielding a branching ratio of 2.99 which is clearly higher but significantly smaller than similar iridate oxide systems measured recently and slightly higher than the pure Iridium metal[43,44]. Laan and Thole proposed that this branching ratio can be used to relate the expectation value of the spin-orbit operator, $<L \cdot S>$ by $BR = (2+r)/(1-r)$ where $r = <L \cdot S>/<n_h>$ and $<n_h>$ is the average number of holes in the 5d level[45]. By this calculation the value of $<L \cdot S>$ was estimated to be 1.32 ($\hbar^2$) which is also small compared to other iridate oxides larger than pure Ir metal. Comparatively smaller value of $<L \cdot S>$ indicates a weaker spin-orbit coupling in the system and probably indicates that it may not be strong enough to lift the degeneracy of the d orbitals (right panel of the Fig. 4c). This is also supported by the single peak feature of the highly symmetric Ir$_{L3}$ white line which should otherwise be an asymmetric double peak feature corresponding to $2p_{3/2} \rightarrow J_{eff} = 5/2$ and $2p_{3/2} \rightarrow J_{eff} = 3/2$ and also essentially indicates the large bandwidth of the 5d valence level in the system which is consistent with our structural and transport analysis showing metallic behavior throughout the temperature range. The electronic properties of pyrochlore iridates are highly sensitive to the bond angle Ir-O-Ir created between corner shared IrO$_6$ octahedra (see Fig. 2e). Clearly, the size of Pb ion, highest among A-site cations of known pyrochlore iridates plays a vital role in determining its bandwidth (W). The difference between the B-O-B bond angle between insulating and metallic pyrochlore iridates has been roughly known to be of the order of 2° [46]. In Pb$_2$Ir$_2$O$_{7-\delta}$ this trend of increasing Ir-O-Ir bond angle coincides with the metallic nature among pyrochlore iridates[47,48], as illustrated in Fig. 1f, the larger Pb cation with its extended 6s orbitals pushes the O ion within the OPb$_2$Ir$_2$ tetrahedra towards the Ir ions making Ir-O-Ir bond angle to be 134°, largest among the reported pyrochlore iridates, resulting in a shorter Ir-O distance and consequently stronger hybridization of extended Ir 5d and O 2p orbitals (see Fig 1f) and higher W. Therefore, experiment proves that this system has a moderate $\lambda$ and also probably a moderate U with a significantly large bandwidth and as a result, metallicity is realized.

However, the temperature dependent resistivity $\rho(T)$ measurement down to 2 K (Fig. 5a) offered a surprise. The resistivity remains linear throughout the temperature range down to atleast 15 K. Most importantly it follows a $\rho \sim T^{3/2}$ dependence below 15 K down to 2 K (see Fig. 5) contrary to the expected $\rho \sim T^2$ variation, expected for a Fermi liquid system at low temperature. Clearly, the system violates the Fermi liquid behavior of a conventional metallic system. However, it's the 4d counterpart pyrochlore does follow a conventional Fermi liquid behaviour[49] of $\rho \sim T^2$. Surprisingly, even at the application of 5 Tesla magnetic field the linear temperature dependence of resistivity does not change except a logarithmic upturn below 15 K and no $T^2$ dependence emerged signifying robust nature of the non-Fermi liquid behavior in the system (Fig. 5c). On the other hand, for a standard metal the maximum resistivity is of the order of $\rho_{max} \sim 1$ $m\Omega cm$[50]. As can be seen from Fig. 5a the resistivity is larger than the $\rho_{max}$ value even at lowest temperature and it does not saturate even at 300 K ($\rho_{300K}$ = 2.4 mΩcm). However, the 4$d$ counter part of the system, $Pb_2Ru_2O_6O'$ tends to saturate at room temperature with a maximum resistivity $\rho_{max} \sim 200$ $\mu\Omega cm$. Evidently $Pb_2Ir_2O_{7-\delta}$ behaves more like a *bad metal* rather than a conventional Fermi liquid metal. Such *bad metals* are often found to violate Mott-Ioffe-Regal (MIR) limit of maximum resistivity[51] ($l \sim a$, $k_F \sim l$, where $l$ is the mean free path, $a$ is the lattice parameter and $k_F$ is the Fermi wave vector) in high $T_c$ cuprate superconductors[14,16], heavy fermions[52] and also recently in some semimetallic pyrochlore iridate systems[20,54] which is indicative of non-Fermi liquid behavior. Violation of MIR often discussed in the context of disordered solids[53]. Therefore, the non-Fermi liquid behavior could arise in this system due to a disordered distribution of $Ir^{4+}$ and $Ir^{5+}$ ions arising due to mixed valent nature of the system which has largely different SOC values unlike in its 4$d$ counterpart $Pb_2Ru_2O_6O'$ as it should be noted that in a solid the strength of spin-orbit coupling is renormalized as $\lambda_{eff} = \lambda_{atomic}/S$ where $S$ is the total spin.

A measure of the heat capacity may give deeper insight into the NFL behavior of the system. Therefore, heat capacity ($C_p$ vs $T$) was measured down to 1.8 K and displayed in Fig. 6a. No $\lambda$ like anomaly was observed in $C_p/T$ vs $T$ data, confirming the absence of any long rage order or any structural phase transition, as confirmed by the XRD analysis of the system. For a Fermi liquid system, low temperature heat capacity ($C_p$) should be linear along with weak $T^3$ lattice term. Therefore, we fitted the low temperature $C_p$ data below 15 K with $C_p = \gamma T + \beta T^3$ where the first part denotes electronic contribution and the later accounts for lattice contribution. However, $C_p/T$ vs $T^2$ strongly deviates below 10 K which is once again consistent with the deviation from quadratic nature of resistivity ($\rho \sim T^2$) of the system. This gives us roughly the $T$ linear component to be $\gamma$ = 1.84 mJ/mol-K$^2$ and $\beta$ = 1.45 mJ/mol-K$^3$. The rather low value of $\gamma$ confirms that there is no perceptible mass enhancement of the fermions in the system. The value of Sommerfeld coefficient signifies it has a very small area on the Fermi surface consistent with the bad metal characteristic as observed from the resistivity measurements as well. On the other hand it could be well fitted at low temperature with $C_p = \gamma'T \ln(T_0/T) + \beta T^3$ below 15 K down to 1.8 K as shown in the inset of Fig. 6b. Logarithmic increase at low temperature is often observed in non-Fermi liquid systems[5,54] and recently it has been also reported in a hole doped pyrochlore system[20]. It might be worth calculating Wilson ratio here which is defined as

$$R_W = \frac{4\pi^2 k_B^2 \chi_0}{(g\mu_B)^2 \gamma} \qquad (1)$$

yielding a value 21.7, which is way larger than the values evidenced for correlated metals with strong spin-orbit coupling ($R_W \sim 1-6$)[55]. This notably high $R_W$ value is an outcome of strong non-Fermi liquid nature of the compound which is consistent with the resistivity analysis. The occurrence of high $R_W$ value in non-Fermi liquids is not unusual and also been observed in the semimetallic NFL iridate $SrIrO_3$ ($R_W \sim 55$)[54] signifying proximity to a quantum critical point.

Now, the magnetic contribution to the heat capacity can be estimated as $C_{mag} = C_p - C_{non-mag}$, where $C_p$ is the total heat capacity. In the absence of suitable non magnetic analogue, the non magnetic contribution can be modelled as $C_{non-mag} = C_{latt} + C_{elec}$. To estimate the lattice contribution, Debye-Einstein equation ($C_{latt} = C(\Theta_{D,E}, T)$) [56] has been used and the electronic contribution is taken as $C_{elec} \sim \gamma T$. However, a low temperature heat capacity fitting may lead to poor estimation of magnetic contribution of the system. Therefore, the total heat capacity $C_p$ was first fitted in the temperature range (55-250 K) and then extrapolated down to 1.8 K (see Fig. 6a) and taken as the $C_{non-mag}$. After subtracting $C_{non-mag}$ from the total $C_p$, finally, the magnetic heat capacity of the sample was estimated. A broad peak was observed in $C_{mag}/T$ vs $T$ around 20 K which is often observed in spin frustrated systems[57–59]. The $C_{mag}$ at low temperature (<10 K) follows an unusual cubic behavior $\sim T^3$ (Fig. 6c). Magnetic entropy of the system was subsequently calculated by the following formula $S_m(T) = \int_0^{T'} C_{mag}/T'$, and the results are displayed in Fig 6d. The magnetic entropy release is 2.6 mJ/mol-Ir-K$^2$ above 40 K which is about 40% of $Rln(2J+1)$ of $J = 1/2$ for $d^5$ Ir ion and almost 25% for the $J = 1$ for $d^4$ Ir ion. The retention of large magnetic entropy at low temperature also indicates highly frustrated nature of the Ir ions in this pyrochlore system which is forbade any long-range magnetic ordering.

To further understand the NFL behavior, we have performed magnetic measurement on the system down to 2 K. Fig. 7 shows the temperature dependence of the dc magnetic susceptibility ($\chi$ vs $T$) down to 2 K, measured in zero field cooled (ZFC) and field cooled(FC) protocols. Having metallic characteristic down to 2 K it was expected to be a Pauli paramagnet like $IrO_2$, however it deviates strongly as a strong upward Curie turn at low temperature is observed. No sign of long-range magnetic order down to 2 K is observed in this nearly featureless paramagnetic like behavior suggesting continuation of spin fluctuation of Ir ions in the system, unlike other $d^5$ pyrochlore iridates[47,61] except $Pr_2Ir_2O_7$, which shows small spin freezing at very low temperature[62]. We have analyzed the susceptibility (In an applied field of 10 kOe) data using the Curie-Weiss (C-W) equation $\chi = \frac{C}{T-\Theta_{CW}} + \chi_0$ ($C$ is the Curie constant while $\Theta_{CW}$ and $\chi_0$ represent the Curie-Weiss temperature and the temperature independent susceptibility, respectively) in the temperature range 100-300 K as shown in Fig. 7a. The fitting yielded a strong negative $\Theta_{CW}$ value of around -96 K and an effective magnetic moment of $\mu_{eff} \sim 0.3$ $\mu_B$/Ir. The large negative $\Theta_{CW}$ of -96 K indicates strong nearest neighbor antiferromagnetic interaction but the absence of any magnetic ordering suggests the presence of inherent geometric frustration in the system expected in a pyrochlore lattice. The closest analogue of the present system in terms of nonmagnetic $A$-site cationic radius, $Bi_2Ir_2O_7$, where Ir is almost entirely present in $4^+$ charge state, also poses metallicity but with strong ferromagnetic instability and the emergence of saturation in magnetic moments vs. field data, unlike the present case. Clearly, while $Bi_2Ir_2O_7$ is on the verge of long-range magnetic order ($\Theta_{CW} \sim -2$ K)[63], $Pb_2Ir_2O_{7-\delta}$ remains far from ordering at least down to 2 K. Therefore, even if pyrochlore iridates are expected to be in $all-in-all-out$ (AIAO) magnetic ground state at low temperature, $Pb_2Ir_2O_{7-\delta}$ becomes another exception in pyrochlore family where frustration seems to dominate. Random distribution of $Ir^{4+}$ and $Ir^{5+}$ ions in the system, which can be thought as a disorder, may also act against any long-range magnetic ordering. On the other hand, the divergence of magnetic susceptibility below 10 K (see Fig. 7b) takes a power law form of $\chi \sim T^{-\alpha}$ ($\alpha \sim 0.21$) which is a typical characteristic of non-Fermi liquids[5], consistent with our transport analysis. Even under the field of 4 T the low temperature $\chi$ retains its power dependence (Fig. 7d) with slightly higher value of $\alpha = 0.23$ compared to the 1 T data, unlike the NFL $SrIrO_3$[54]. Although the $SrIrO_3$ has been found to reside near a quantum critical point (QCP) ($T = 0$ and $\mu_0 H = 0.23$ T) where the low temperature magnetic susceptibility exponent is highly sensitive to very low applied magnetic fields indicating a ferromagnetic instability in the system, this is not the case for the present system. Therefore, $Pb_2Ir_2O_{7-\delta}$ shows the robustness of the NFL characteristic and the quantum critical point could not be realized down to 2 K and at high field. One should note that $SrIrO_3$ is purely stoichiometric compound whereas $Pb_2Ir_2O_{7-\delta}$ has oxygen off-stoichiometry leading to distribution of $Ir^{4+}$ and $Ir^{5+}$ in the system.

To get more insight of the spin dynamics, local magnetic probe muon spin relaxation has been performed. The temperature dependent $\mu$SR spectra between 80 mK and 30 K at zero field is shown in Fig. 8. However, no signature of oscillation was observed down to at least 80 mK confirming the absence of long-range magnetic order which implies the preservation of the time reversal symmetry in the system. In contrary to the anticipated AIAO magnetic configuration where all of the spins of magnetic ions pointing either inward or outward with respect to the center of the tetrahedra[64,65], disordered distribution between $Ir^{4+}$ and $Ir^{5+}$ may hinder the long-range magnetic order in the present system. In the hole doped pyrochlore iridate system $Eu_{2-x}Sr_xIr_2O_7$ [20] NFL state was observed with the simultaneous suppression of magnetic order. This is in line with theoretical claim by Eun-Gook Moon *et. al.*[19] which predicts non-Fermi liquid state in pyrochlore iridates provided time-reversal and cubic symmetries are protected.

One can also calculate the frustration parameter $f_N = \Theta_{CW}/T_N$, ($T_N$ is the lowest probed temperature), from $C - W$ fit in the magnetic susceptibility measurement which comes out to be as large as 1200. Nonetheless, the extreme slow relaxation nature of muon spin is somewhat surprising given the strong geometric frustration in the system. The relaxation does not evolve with temperature, as the signal remains almost similar between 80 mK to 30 K. This signifies the complete absence of freezing of Ir moments down to atleast 80 mK. Evidently the system stays in a fast-fluctuating paramagnetic limit in the whole temperature limit. Upon application of mere 20 Oe longitudinal field the feeble relaxation completely dies off. The weak decoupling nature of muon spins signify the nature of relaxation is static originating from nuclei moments[67]. However, one should take note that only 20% of Pb bears a nuclear spin[66] and the one on Ir is super small. On the other hand, the muons stop far from these 20% Pb nuclear spin near Ir moments which are very small. Clearly this weak relaxation arises due to weak dipolar Pb nuclear interaction and small Ir moments.

### IV. SUMMARY AND CONCLUSION

In summary we have demonstrated that $Pb_2Ir_2O_{7-\delta}$ is a weakly correlated metal with moderate SOC and mixed Ir valency that exhibits unusual electronic transport, magnetic and thermodynamics properties that deviate from the normal Fermi liquid behaviors. The linear temperature dependence in resistivity over a large temperature range and typical $T^{3/2}$ dependence below 15 K is observed by experiment. On the other hand, magnetic contribution at low $T$ follows a power law behavior which is retained even under a strong field indicates onset of a non-Fermi liquid state in the system. The $\mu$SR and magnetic susceptibility data discarded any sign of long-range magnetic ordering in the system down to at least 80 mK. The random distribution between $Ir^{4+}$ and $Ir^{5+}$ which can be thought as disorder may prevent the system from being ordered. Strong inherent geometric frustration (frustration parameter $f_N \sim 1200$) is also prevalent in the system. Despite having metallicity down to the lowest measured temperature, it shows very small Sommerfeld coefficient in the low temperature heat capacity fitting, signifying small density of low energy excitations at Fermi level, and consequently a large Wilson ratio ($R_W \sim 22$) asserting its robust non-Fermi liquid nature. Absence of magnetic ion at $A$ site and a very low value of Sommerfeld coefficient ($T$ linear term in heat capacity), strongly discards Kondo mechanism to be responsible for the observed NFL nature in the system. Also, there is no sign of Kondo resistivity in the sample. Our analysis does not find the proximity to quantum critical point (QCP). However, a pressure dependent study might be very useful for deeper understanding of this new paradigm of NFL phenomenon. The non-Fermi liquid behavior in $Pb_2Ir_2O_{7-\delta}$ could be ascribed to the disordered distribution of $Ir^{4+}$ and $Ir^{5+}$ which has different strengths of spin-orbit coupling. Therefore, locally there could be largely different coupling due to the distribution of $d^4$ ($S = 1$) and $d^5$ ($S=1/2$) ions. Nonetheless, the comparatively new NFL phenomenon in pyrochlore iridates adds to the myriad of complex electronic and magnetic ground state it hosts.

## V. ACKNOWLEDGEMENT

MSK thanks UGC India and IACS for fellowship. SR thank Technical Research Center (TRC) of IACS for providing experimental facilities. SR also thanks SERB, DST for financial support (Project no. CRG/2019/003522).

---

TABLE I. Rietveld refined crystallographic parameters of $Pb_2Ir_2O_{7-\delta}$. (a) (300 K): $a$ = 10.289(1) Å, space group: $Fd\bar{3}m$, $O_x$ = 0.325 $R_p$ = 12.6, $R_{wp}$= 15.5, $R_{exp}$= 8.28 , and $\chi^2$=3.49. (b) (4 K): $a$ = 10.266(7) Å, space group: $Fd\bar{3}m$, $O_x$ = 0.328, $R_p$ = 14.3, $R_{wp}$= 17.5, $R_{exp}$= 8.40, and $\chi^2$=4.32

| | | $Pb_2Ir_2O_{7-\delta}$ | | | | |
|---|---|---|---|---|---|---|
| Temperature | Atoms | occupancy | x | y | z | B ($\times 10^{-3}$Å$^2$) |
| 300 K | Pb | 1.0 | 0.5 | 0.5 | 0.5) | 2.3(7) |
| | Ir | 1.0 | 0 | 0 | 0 | 0.9(5) |
| | O | 1.0 | 0.325(3) | 0.125 | 0.125 | 1.2(1) |
| | O´ | 0.72(5) | 0.375 | 0.375 | 0.375 | 4.7(6) |
| 4 K | Pb | 1.0 | 0.5 | 0.5 | 0.5) | 4.7(6) |
| | Ir | 1.0 | 0 | 0 | 0 | 4.7(6) |
| | O | 1 | 0.328(1) | 0.325 | 0.325 | 1.8(5) |
| | O´ | 0.76(4) | 0.375 | 0.375 | 0.375 | 1.8(5) |

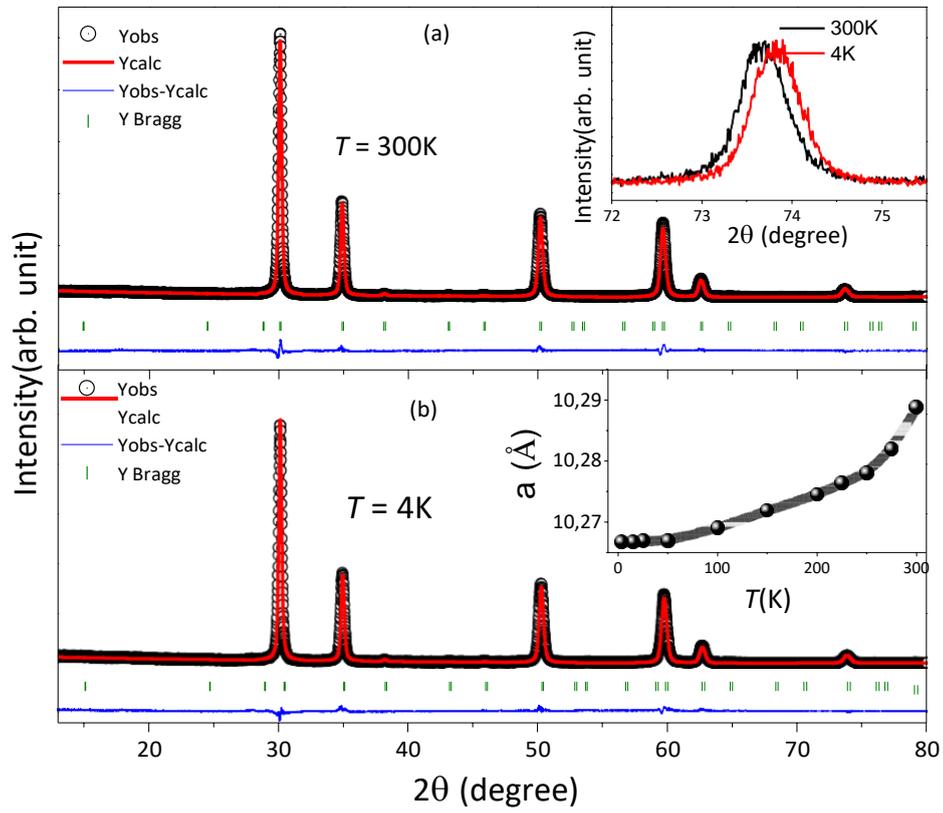

FIG. 1. (color online) (a) Rietveld refined XRD pattern of $Pb_2Ir_2O_{7-\delta}$ sample at (a) 4 K (inset: an enlarged view of the XRD pattern at a higher angle between 300 K and 4 K data) (b) at 300 K (inset: temperature variation of lattice parameter $a$ (Å))

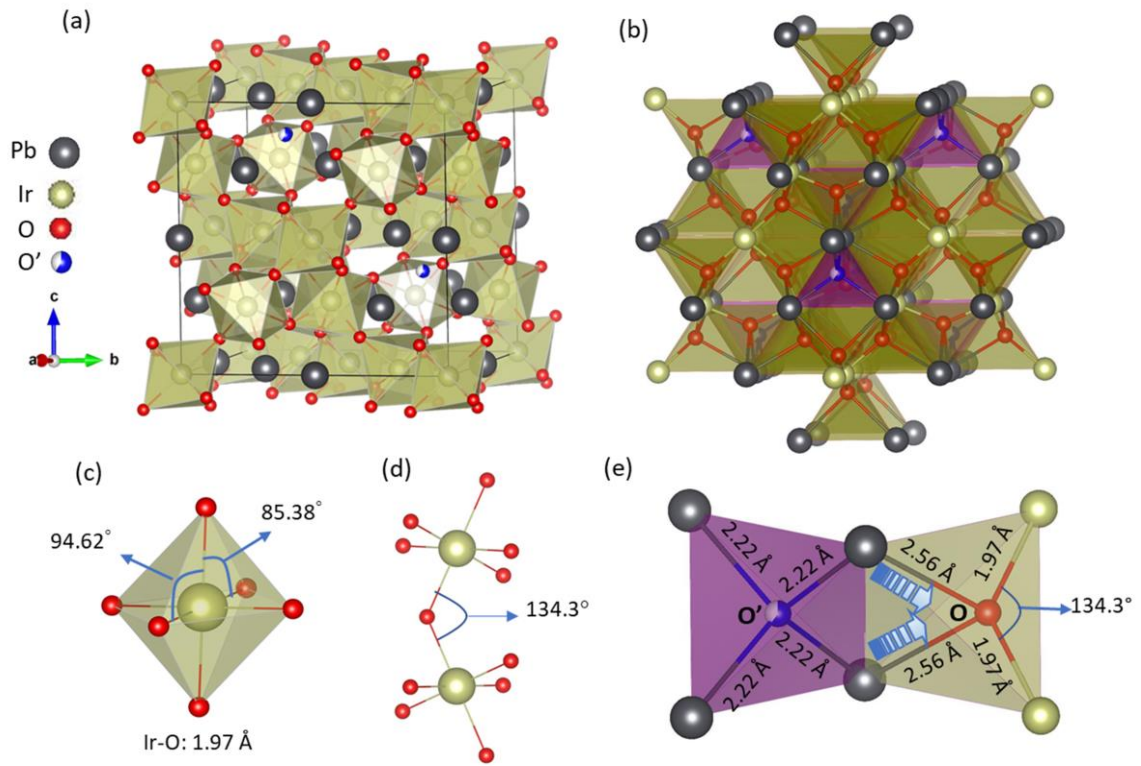

FIG. 2. (Color online) (a) Rietveld refined crystal structure showing $IrO_6$ corner shared octahedral network, (b) edge shared tetrahedral network between O´ $Pb_4$ and $OPb_2Ir_2$, (c) single $IrO_6$ distorted octahedra, (d) Ir-O-Ir bond angle between two corner shared $IrO_6$ octahedra, (e) a single edge shared tetrahedral unit of O´ $Pb_4$ and $OPb_2Ir_2$ with bond length and bond angles

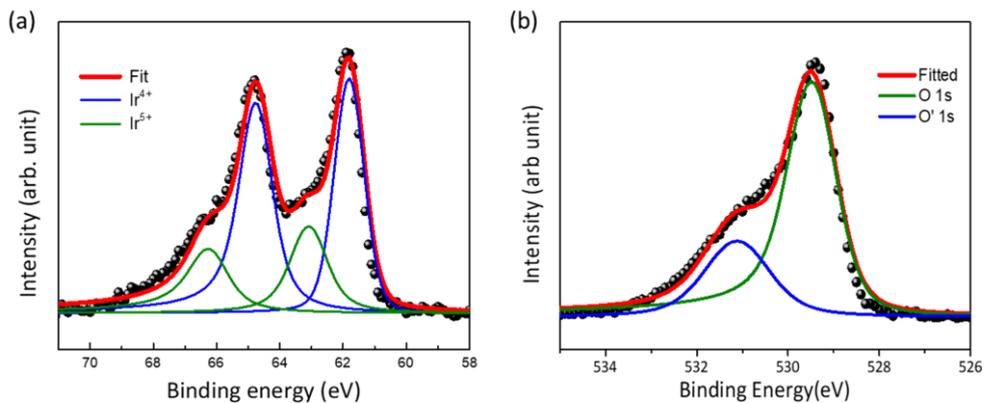

FIG. 3. (Color online) (a) Ir *4f* core level XPS spectrum (shaded black circles) along with the fitting (red solid line), showing the contribution of $Ir^{5+}$ (green) and $Ir^{4+}$ (blue) respectively. (b) Oxygen 1*s* spectra showing contribution from O and O´ 1*s* level

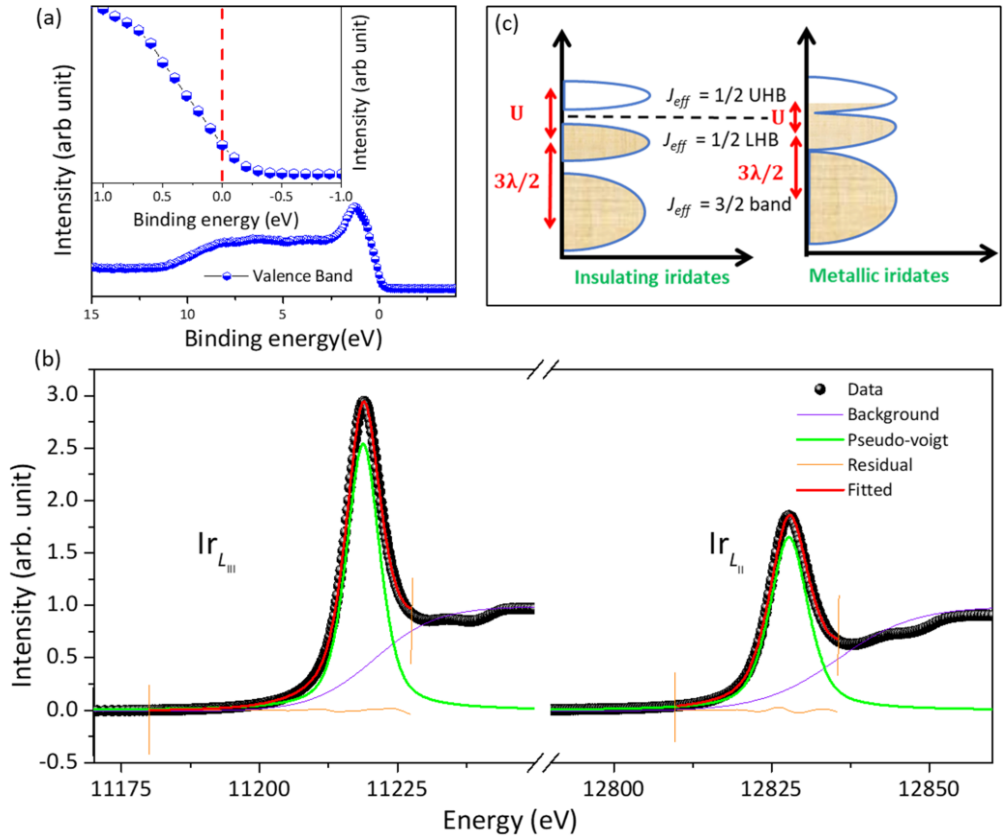

FIG. 4. (Color online) (a) XPS valence band spectra of $Pb_2Ir_2O_{7-\delta}$ system, inset: zoomed view of spectra near Fermi level, (b) Normalized $Ir_{L3}$ and $Ir_{L2}$ edge white line spectra (shaded black circles) along with the fitting (red solid line), (c) Schematic diagram of Ir 5$d$ level splitting in presence of $U$ and SOC.

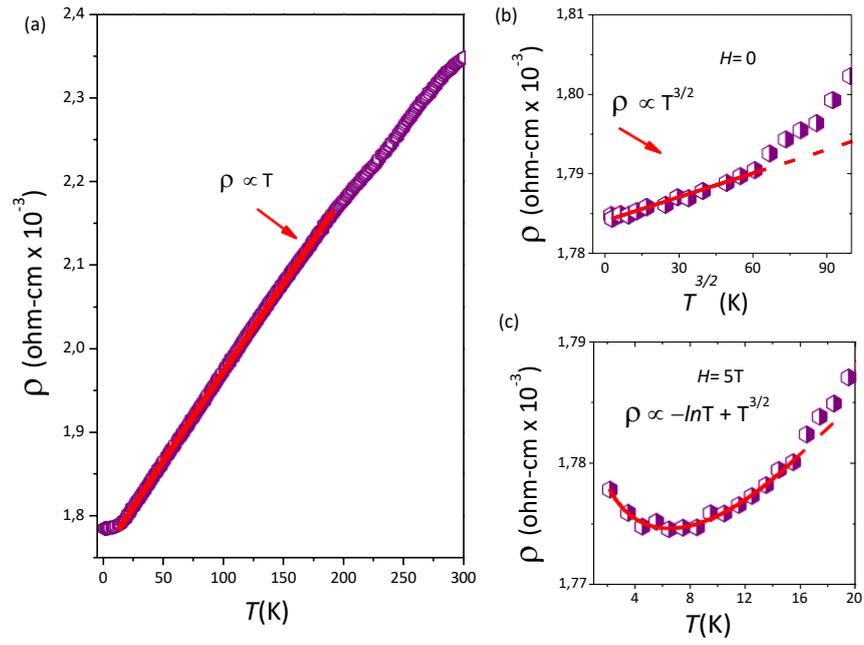

FIG. 5. (Color online) (a) Linear fitting of resistivity data below 200 K down to 15 K. (b) Low Temperature fitting of zero field resistivity data. (c) Low Temperature fitting of 5 T field resistivity data

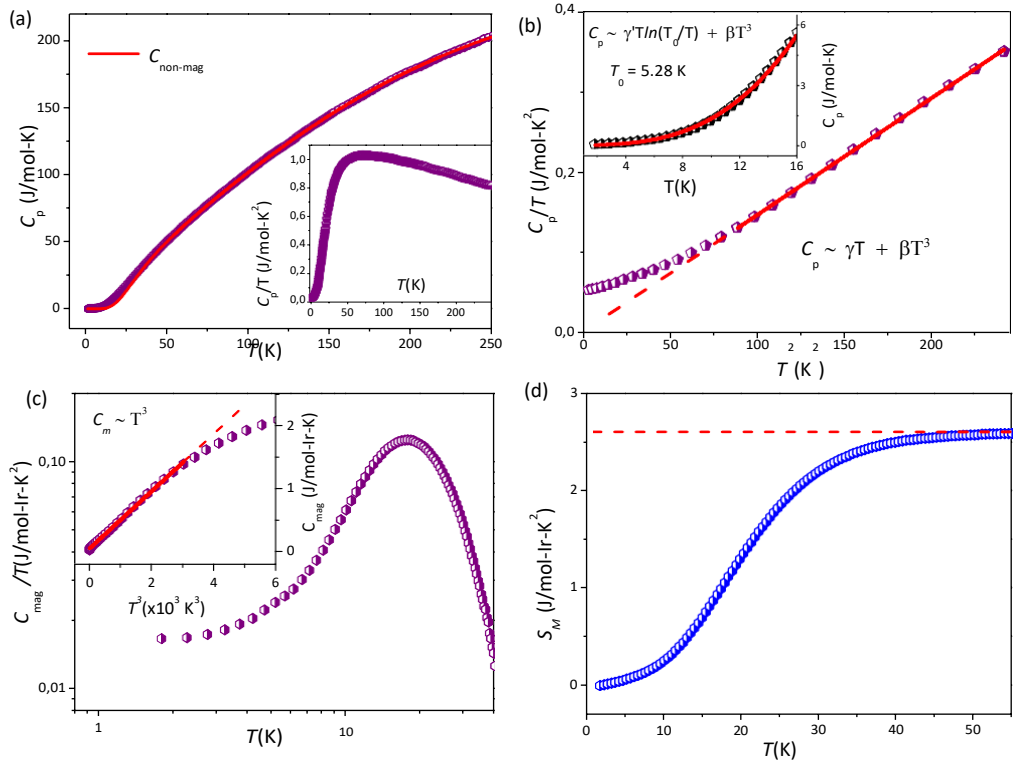

FIG. 6. (color online)(a)Temperature dependence of total specific heat $C_p$ (purple sphere)in whole temperature range along with Debye-Einstein(DE) fitting (red solid line) for lattice contribution at zero field, inset: $C_p/T$ vs $T$ plot in the same temperature range, , (b) Deviation of $C_p=\gamma T +\beta T^3$ nature at low $T$ in $C_p/T$ vs $T$ data ; in the inset low temperature fitting of $C_p$ vs $T$ data with modified $C_p=\gamma'T \ln(T_0/T )+ \beta T^3$ ,(c) Magnetic contribution $C_{mag}$/T vs $T$ in logarithmic scale; in the inset power low dependence of $C_{mag}$ vs $T$ at low Temperature (<10 K), (d) Magnetic entropy($S_M$) release as a function of temperature (dashed red line shows the maximum entropy release)

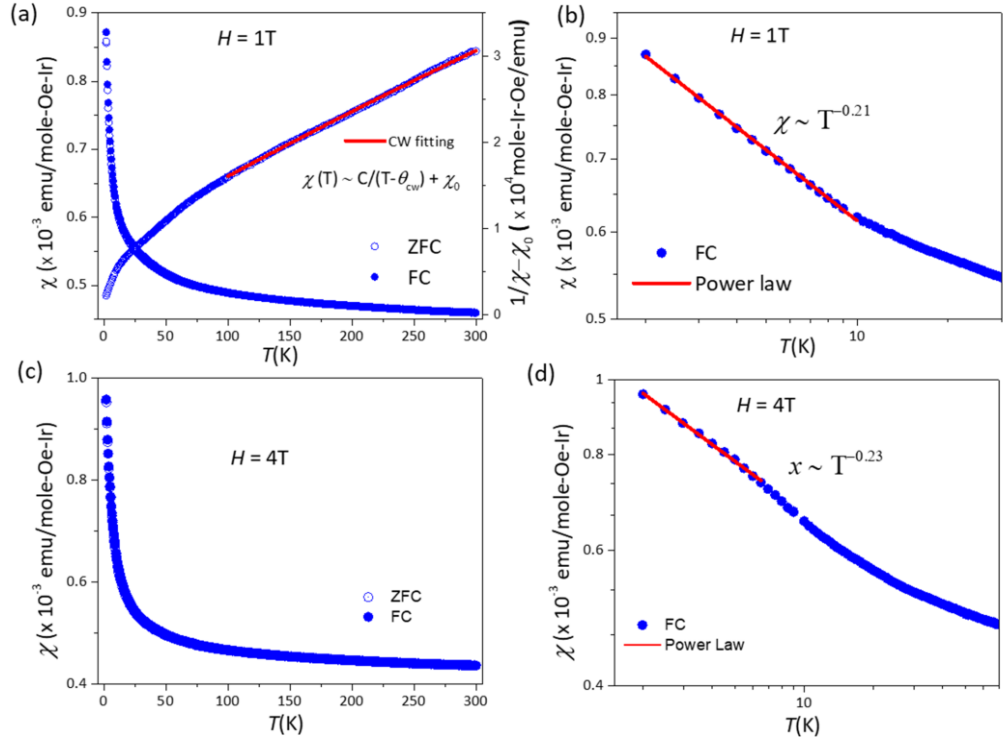

FIG. 7. (Color online) (a) Temperature dependent dc susceptibility variations at 1 T during zero-field-cooled (open blue circles) and field-cooled (shaded blue circles) protocols, Right panel; Temperature dependence of $1/(\chi-\chi_0)$ is and Curie-Weiss fitting (red solid line) plotted, (b) Log-log plot of low temperature magnetic susceptibility data at 1 T (blue solid circle) and fitted curve (red solid line). (c) Temperature dependent dc susceptibility variations at 4 T during zero-field-cooled (open blue circles) and field-cooled (shaded blue circles) protocols, (d) Log-log plot of low temperature magnetic susceptibility data at 4 T (blue solid circle) and fitted curve (red solid line)
25

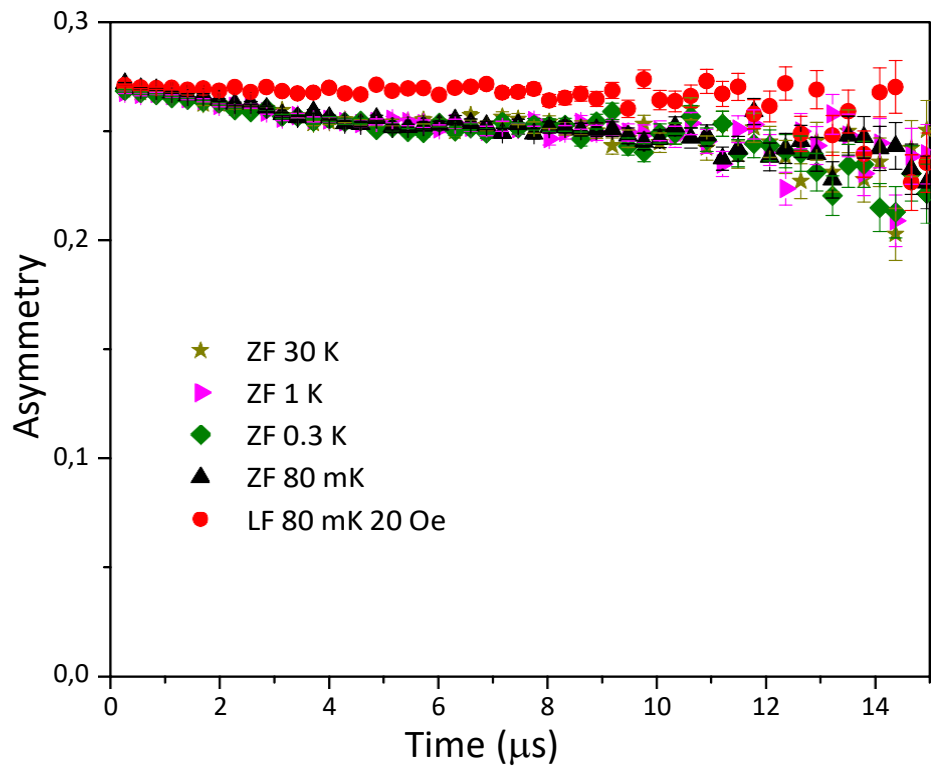

FIG. 8. (Color online) Time evolution of zero field $\mu$SR spectra between 80 mK and 30 K and longitudinal field (20 Oe) spectra at 80 mK